\documentclass[aps,pra,showpacs,floatfix,twocolumn]{revtex4-1}
\usepackage{graphicx}
\usepackage{amsmath}
\usepackage{bm}
\usepackage{color}
\usepackage{dcolumn}
\usepackage{hyperref}

\begin{document}


\newcommand{\etal}{{\it et al.}}
\newcommand{\fref}[1]{Fig.~\ref{#1}}
\newcommand{\Fref}[1]{Figure \ref{#1}}
\newcommand{\sref}[1]{Sec.~\ref{#1}}
\newcommand{\Eref}[1]{Eq.~(\ref{#1})}
\newcommand{\tref}[1]{Table~\ref{#1}}
\newcommand{\rtw}{\longrightarrow}
\newcommand{\veps}{\varepsilon}
\newcommand{\cm}{cm$^{-1}$}
\newcommand{\ignoretext}[1]{}
\newcommand{\fracline}[2]{#1/#2}

\newcommand{\cmt}[1]{\textbf{[\![}#1\textbf{]\!]}}
\newcommand{\pprime}{{\prime\prime}}
\newcommand{\s}{\ensuremath{^1\!S_0}}
\newcommand{\p}{\ensuremath{^3\!P_0}}
\newcommand{\pa}{\ensuremath{^3\!P_1}}
\newcommand{\pb}{\ensuremath{^1\!P_1}}
\newcommand{\yb}[1]{$^{#1}$Yb}
\newcommand{\sr}[1]{$^{#1}$Sr}
\newcommand{\hg}[1]{$^{#1}$Hg}

\newcommand{\thg}{$^{229g}$Th}
\newcommand{\thm}{$^{229m}$Th}
\newcommand{\thgion}{$^{229g}$Th$^{3+}$}
\newcommand{\thmion}{$^{229m}$Th$^{3+}$}


\newcommand{\NIST}{
National Institute of Standards and Technology, 325 Broadway, Boulder, Colorado 80305, USA}





\title{Hyperfine structure in \thgion{} as a probe of the \thg\,$\rightarrow$\,\thm{} nuclear excitation energy
}

\author{K. Beloy}
\affiliation{\NIST}

\date{\today}

\begin{abstract}
We identify a potential means to extract the \thg\,$\rightarrow$\,\thm{} nuclear excitation energy from precision microwave spectroscopy of the $5F_{5/2,7/2}$ hyperfine manifolds in the ion \thgion{}.
The hyperfine interaction mixes this ground fine structure doublet with states of the nuclear isomer, introducing small but observable shifts to the hyperfine sub-levels. We demonstrate how accurate atomic structure calculations may be combined with measurement of the hyperfine intervals to quantify the effects of this mixing. Further knowledge of the magnetic dipole decay rate of the isomer, as recently reported, allows an indirect determination of the nuclear excitation energy.
\end{abstract}


\pacs{21.10.-k, 32.10.Fn}
\maketitle


Thorium-229 is extraordinary among nuclei in that it possesses an isomer state \thm{} lying within several eV of its ground state \thg{}.
The anomalously small nuclear excitation frequency is predicted to be within range of modern lasers, opening the door to a number of scientific possibilities. Proposed clocks based on this transition hold promise for unprecedented metrological performance, as the compact nuclear charge distribution (relative to the electron cloud in an atom) renders the ultra-narrow transition frequency largely insusceptible to environmental influences \cite{PeiTam03,CamSteChu09,RelDeMGre10,CamRadKuz12,KazLitRom12}. 
Moreover, such a nuclear clock could be a valuable instrument for testing stability of fundamental constants, including the fine structure constant $\alpha$, as the ``accidental'' near-degeneracy of the two nuclear states (relative to typical nuclear energy scales) renders the transition frequency highly-sensitive to fundamental constant variation \cite{Fla06}.
Further still, it has been suggested that this transition could be used to realize a novel nuclear-based laser \cite{Tka11}.




Presently, the nuclear excitation energy $\Delta_\mathrm{nuc}$ has only been determined indirectly through differencing schemes, using
$\gamma$-radiation observed following $\alpha$-decay of $^{233}$U into $^{229}$Th.
More than two decades ago, Reich and Helmer \cite{ReiHel90,HelRei94} deduced the result $-1\pm4$ eV, following this a few years later with the refined value $3.5\pm1$ eV. 
After another decade, Barci \etal~\cite{BarArdBar03} and Guimar\~{a}es-Filho and Helene \cite{GuiHel05} reported the results $3.4\pm1.8$ eV and $5.5\pm1$ eV, respectively.
Finally, Beck~\etal~\cite{BecBecBei07,BecWuBei10} presented the result $7.6\pm0.5$ eV, which they later modified slightly to $7.8\pm0.5$ eV.
While the most recent value of Beck \etal~is now largely accepted by the community, the large discrepancy with earlier results is not well-understood.
In a critique of the above works, however, Sakharov \cite{Sak10} suggested that the literature values suffer from systematic errors and underestimated uncertainties, further clouding precise knowledge of $\Delta_\mathrm{nuc}$. 
New, independent means of determining $\Delta_\mathrm{nuc}$ could prove invaluable for efforts towards
laser-excitation of the nucleus
\cite{CamSteChu09,CamRadKuz11,RelDeMGre10,HehGreRel13}, either by providing improved results or through corroboration or dismissal of existing literature values.

\begin{figure}[h!]
\includegraphics[width=\linewidth]{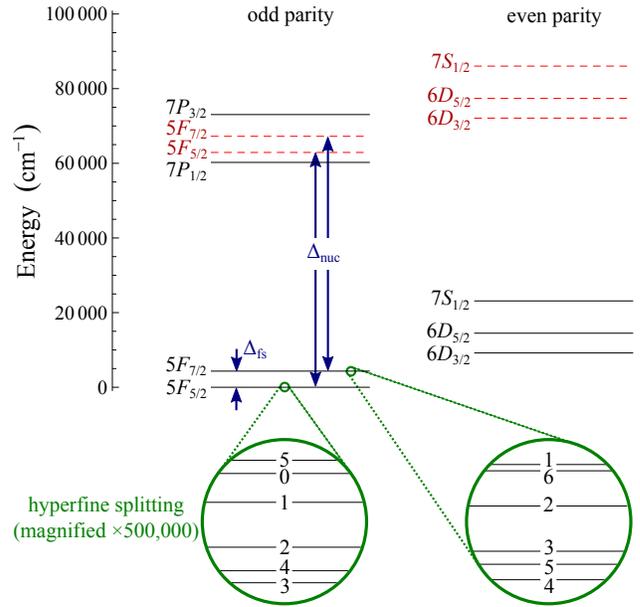}
\caption{(color online) Energy levels of $^{229}$Th$^{3+}$, including nuclear and electronic excitations. Solid-black and red-dashed lines distinguish levels associated with the ground and isomer nuclear states, \thg{} ($I^\pi\!=\!\frac{5}{2}^+$) and \thm{} ($I^\pi\!=\!\frac{3}{2}^+$), respectively. Hyperfine manifolds of the ground fine structure doublet are magnified, with sub-levels labeled according to total angular momentum $F$, where $|I\!-\!J|\leq F\leq I\!+\!J$. Hyperfine mixing between like-parity states has small but observable effects on these hyperfine manifolds, with the magnitude of mixing being dependent on level separation (e.g., $\Delta_\mathrm{fs}$ and $\Delta_\mathrm{nuc}$). For the purpose of illustration, a nuclear excitation energy of $\Delta_\mathrm{nuc}=7.8$ eV is assumed here.
}
\label{Fig:ladder}
\end{figure}

In this paper we identify a potential means to extract $\Delta_\mathrm{nuc}$ from precision microwave spectroscopy of the $5F_{5/2,7/2}$ hyperfine manifolds in the ion \thgion{} (see Fig.~\ref{Fig:ladder}). The proposed method further relies on capabilities of state-of-the-art atomic structure calculations for this system~\cite{SafSafRad13} as well as knowledge of the \thm{}\,$\rightarrow$\,\thg{} magnetic dipole nuclear decay rate~\cite{ZhaEscNat12}. Campbell \etal~\cite{CamRadKuz11} have demonstrated laser cooling of \thgion{} within a linear Paul trap and, furthermore, spectroscopically resolved $5F_{5/2,7/2}$ and $6D_{3/2,5/2}$ hyperfine sub-levels. Their measurement precision was sufficient to deduce hyperfine constants $A$ and $B$ for all four states, though the optical spectroscopy employed ($5F_{5/2,7/2}\rightarrow 6D_{3/2,5/2}$) did not fully benefit from the long-lived nature of the $5F_{5/2,7/2}$ states, which makes the hyperfine intervals of these two states amenable to very high-precision microwave spectroscopy. In Ref.~\cite{SafSafRad13}, it is argued that such measurements would not only be capable of yielding much-improved $A$ and $B$ constants, but also of revealing additional constants in the hierarchy of hyperfine constants. 
In principle, five constants are required to fully characterize the five intervals in each hyperfine manifold (refer to Fig.~\ref{Fig:ladder}).

The hyperfine interaction accounts for electromagnetic coupling of atomic electrons with the nucleus, beyond that of the dominant electric monopole (Coulomb) interaction. 
Decomposed into multipolar contributions, the hyperfine interaction reads $V_\mathrm{hfi}=\sum_k{\mathcal{M}^{(k)}\cdot\mathcal{T}^{(k)}}$, where $\mathcal{M}^{(k)}$ and $\mathcal{T}^{(k)}$ are rank-$k$ operators acting in the nuclear and electronic subspaces, respectively, and the intervening dot signifies a scalar product. Expressions for $\mathcal{M}^{(k)}$ and $\mathcal{T}^{(k)}$ may be found, for example, in Refs.~\cite{Sch55,Joh07,BelDerJoh08,PorFla10}. The summation runs over positive integers $k$, with terms ascending in $k$ describing magnetic dipole ($M1$), electric quadrupole ($E2$), magnetic octupole ($M3$), electric hexadecapole ($E4$), etc., multipolar interactions. Hyperfine constants $A$, $B$, $C$, $D$, etc., quantify the effect of the respective interactions to first order.
Following conventional definitions \cite{Sch55,BelDerJoh08}, hyperfine constants of the $5F_{5/2,7/2}$ states in \thgion{} read 
\begin{equation}
\begin{gathered}
A_{5/2}\equiv
\frac{2}{105}\,
\mathcal{M}^{(1)}_{\downarrow\downarrow}
\mathcal{T}^{(1)}_{\downarrow\downarrow};
\qquad
A_{7/2}\equiv
\frac{1}{21\sqrt{15}}\,
\mathcal{M}^{(1)}_{\downarrow\downarrow}
\mathcal{T}^{(1)}_{\uparrow\uparrow},
\\
B_{5/2}\equiv
\frac{5}{21}\,
\mathcal{M}^{(2)}_{\downarrow\downarrow}
\mathcal{T}^{(2)}_{\downarrow\downarrow};
\qquad
B_{7/2}\equiv
\frac{1}{3\sqrt{2}}\,
\mathcal{M}^{(2)}_{\downarrow\downarrow}
\mathcal{T}^{(2)}_{\uparrow\uparrow},
\\
C_{5/2}\equiv
\frac{5}{252}\,
\mathcal{M}^{(3)}_{\downarrow\downarrow}
\mathcal{T}^{(3)}_{\downarrow\downarrow};
\qquad
C_{7/2}\equiv
\frac{1}{12}\sqrt{\frac{5}{66}}\,
\mathcal{M}^{(3)}_{\downarrow\downarrow}
\mathcal{T}^{(3)}_{\uparrow\uparrow},
\\
D_{5/2}\equiv
\frac{1}{252}\,
\mathcal{M}^{(4)}_{\downarrow\downarrow}
\mathcal{T}^{(4)}_{\downarrow\downarrow};
\qquad
D_{7/2}\equiv
\frac{1}{36\sqrt{22}}\,
\mathcal{M}^{(4)}_{\downarrow\downarrow}
\mathcal{T}^{(4)}_{\uparrow\uparrow},
\end{gathered}
\label{Eq:ABCD}
\end{equation}
with
$\mathcal{M}^{(k)}_{ij}\equiv\langle i||\mathcal{M}^{(k)}||j\rangle$
and
$\mathcal{T}^{(k)}_{ij}\equiv\langle i||\mathcal{T}^{(k)}||j\rangle$
being reduced matrix elements and where the associations ${\downarrow,\!\uparrow}={^{229g,m}\mathrm{Th}}$ and ${\downarrow,\!\uparrow}=5F_{5/2,7/2}$ are introduced for the two subspaces. Diagonal nuclear matrix elements appearing here are proportional to 
magnetic dipole $\mu\propto\mathcal{M}^{(1)}_{\downarrow\downarrow}$, electric quadrupole $Q\propto\mathcal{M}^{(2)}_{\downarrow\downarrow}$, magnetic octupole $\Omega\propto\mathcal{M}^{(3)}_{\downarrow\downarrow}$, and electric hexadecapole $\Pi\propto\mathcal{M}^{(4)}_{\downarrow\downarrow}$ 
moments of the \thg{} nucleus.


In principle, hyperfine constants may be determined spectroscopically by taking appropriate linear combinations of measured hyperfine energy intervals.
However, spectroscopy does not differentiate between first order effects of the hyperfine interaction and all higher order effects, and for high precision measurements it becomes necessary to distinguish {\it spectroscopic}, or ``uncorrected,'' hyperfine constants from lowest order, or ``corrected,'' hyperfine constants.
While this distinction is only relevant at the ppm level for $A$ and $B$ constants, it becomes essential for the $C$ and $D$ constants, as second order $M1$-$M1$, $M1$-$E2$, or $E2$-$E2$ shifts to hyperfine sub-levels may be comparable to the first order $M3$ or $E4$ shifts.
In the remainder, a tilde is used to distinguish spectroscopic hyperfine constants $\widetilde{A}$, $\widetilde{B}$, $\widetilde{C}$, and $\widetilde{D}$, as determined from the hyperfine intervals, 
from their lowest order counterparts, given by Eq.~(\ref{Eq:ABCD}).

We illustrate the influence of higher order effects by initially focusing on the $\widetilde{D}$ constants, writing each as the sum of the three terms
\begin{equation}
\widetilde{D}_J=D_J+D^\prime_J+d^\prime_J,
\label{Eq:D3term}
\end{equation}
where $D_J$ is given by Eq.~(\ref{Eq:ABCD}), $D^\prime_J$ includes dominant second order contributions, and $d^\prime_J$ subsumes all remaining higher order contributions. General angular considerations prohibit second order $M1$-$M1$ and $M1$-$E2$ effects from entering the $\widetilde{D}$ constants \cite{Sch55}, limiting the $D^\prime_J$ here to $E2$-$E2$ contributions. For the $5F_J$ state, we explicitly consider contributions attributed to mixing with i) the neighboring $5F_{J^\prime}$ state, ii) the $5F_J$ state of the isomer, and iii) the $5F_{J^\prime}$ state of the isomer.
The three contributions are proportional to $\Delta_\mathrm{fs}^{-1}$, $\Delta_\mathrm{nuc}^{-1}$, and $(\Delta_\mathrm{nuc}\pm\Delta_\mathrm{fs})^{-1}$, respectively, where $\Delta_\mathrm{fs}$ is the fine structure splitting. For the third term, we take $(\Delta_\mathrm{nuc}\pm\Delta_\mathrm{fs})^{-1}
\rightarrow\Delta_\mathrm{nuc}^{-1}$, as is valid in the limit $\Delta_\mathrm{fs}\ll\Delta_\mathrm{nuc}$; formally, the omitted part in this substitution is absorbed by the residual term $d_J^\prime$. From a second order analysis, the contributions read
\begin{equation*}
\begin{gathered}
D_{5/2}^\prime\equiv
\frac{\left|
\mathcal{M}^{(2)}_{\downarrow\downarrow}
\mathcal{T}^{(2)}_{\downarrow\uparrow}
\right|^2}
{20580\,\Delta_\mathrm{fs}}
+
\frac{\left|
\mathcal{M}^{(2)}_{\downarrow\uparrow}
\mathcal{T}^{(2)}_{\downarrow\downarrow}
\right|^2}
{3430\,\Delta_\mathrm{nuc}}
-
\frac{4\left|
\mathcal{M}^{(2)}_{\downarrow\uparrow}
\mathcal{T}^{(2)}_{\downarrow\uparrow}
\right|^2}
{46305\Delta_\mathrm{nuc}},
\\
D_{7/2}^\prime\equiv
-
\frac{\left|
\mathcal{M}^{(2)}_{\downarrow\downarrow}
\mathcal{T}^{(2)}_{\downarrow\uparrow}
\right|^2}
{2940\,\Delta_\mathrm{fs}}
+
\frac{\left|
\mathcal{M}^{(2)}_{\downarrow\uparrow}
\mathcal{T}^{(2)}_{\uparrow\uparrow}
\right|^2}
{2205\,\Delta_\mathrm{nuc}}
-
\frac{4\left|
\mathcal{M}^{(2)}_{\downarrow\uparrow}
\mathcal{T}^{(2)}_{\downarrow\uparrow}
\right|^2}
{6615\Delta_\mathrm{nuc}}.
\end{gathered}
\end{equation*}
Introducing the off-diagonal hyperfine constant $B_\mathrm{o.d.}\equiv (5/36)\mathcal{M}^{(2)}_{\downarrow\downarrow}
\mathcal{T}^{(2)}_{\downarrow\uparrow}$ and the dimensionless parameter
\begin{equation}
\eta_k\equiv
\frac{
\mathcal{M}^{(k)}_{\downarrow\uparrow}
/\mathcal{M}^{(k)}_{\downarrow\downarrow}
}
{\sqrt{\Delta_\mathrm{nuc}/\Delta_\mathrm{fs}}},
\label{Eq:etak}
\end{equation}
these expressions may be recast as
\begin{equation}
\begin{gathered}
D_{5/2}^\prime=
\frac{
3\left[
72\,B_\mathrm{o.d.}^2
+
147\,\eta_2^2\,B_{5/2}^2
-
128\,\eta_2^2\,B_\mathrm{o.d.}^2
\right]
}
{85750\,\Delta_\mathrm{fs}},
\\
D_{7/2}^\prime=
\frac{
2\left[
-54\,B_\mathrm{o.d.}^2
+
25\,\eta_2^2\,B_{7/2}^2
-
96\,\eta_2^2\,B_\mathrm{o.d.}^2
\right]
}
{6125\,\Delta_\mathrm{fs}}.
\end{gathered}
\label{Eq:Drecast}
\end{equation}
Note that the influence of the isomer state is contained within the parameter $\eta_2$.

\begin{table}[t!]
\caption{
Nuclear and electronic properties contributing to spectroscopic (i.e., uncorrected) hyperfine constants $\widetilde{C}_{5/2}$, $\widetilde{C}_{7/2}$, $\widetilde{D}_{5/2}$, and $\widetilde{D}_{7/2}$. Many entries should be regarded as estimates only. Here $e$ is the elementary charge, $\mu_N$ the nuclear magneton, and b the barn unit of area. Reference ``p'' denotes present {\it ab initio} theoretical electronic matrix elements (see text). Respective values of $\Delta_\mathrm{nuc}$ correspond to 3.5(10), 3.4(18), 5.5(10), and 7.8(5) in units of eV.
}
\label{Tab:props}
\begin{ruledtabular}
\begin{tabular}{llcc}
\multicolumn{2}{l}{Property (unit)} & Values & Refs. \\
\hline
\vspace{-3mm}\\
\multicolumn{4}{c}{nuclear properties}\\
$\mathcal{M}_1^{\downarrow\downarrow}$ & $(\mu_N)$
			& 1.3(1), 1.04(2)	& 
			[\onlinecite{GerLucVer74},\onlinecite{SafSafRad13}] \\
$\mathcal{M}_2^{\downarrow\downarrow}$ & $(e\mathrm{b})$
			& 6.45\footnotemark[1], 6.4(1)	& 
			[\onlinecite{BemMcGFor88},\onlinecite{SafSafRad13}] \\
$\mathcal{M}_3^{\downarrow\downarrow}$ & $(\mu_N\mathrm{b})$
			& 0.43\footnotemark[2]		&  \\
$\mathcal{M}_4^{\downarrow\downarrow}$ & $(e\mathrm{b}^2)$	
			& 1.4\footnotemark[1]		& \cite{BemMcGFor88} \\
$|\mathcal{M}_1^{\downarrow\uparrow}|$\footnotemark[3] & $(\mu_N)$
			& 1.2, 0.85, 0.65		& 
[\onlinecite{DykTka98},\onlinecite{BarArdBar03},\onlinecite{RucPloZyl06}]\\
$\mathcal{M}_2^{\downarrow\uparrow}$ & $(e\mathrm{b})$
			& 0.80, 2.4		& 
[\onlinecite{BarArdBar03},\onlinecite{RucPloZyl06}] \\
$\Delta_\mathrm{nuc}$ & \multicolumn{2}{l}{$(10^{15}~\mathrm{Hz})$
~ 0.8(2), 0.8(4), 1.3(2), 1.9(1)}& 
[\onlinecite{HelRei94},\onlinecite{BarArdBar03},\onlinecite{GuiHel05},\onlinecite{BecWuBei10}] \\
\vspace{-3mm}\\
\multicolumn{4}{c}{electronic properties}\\
$\mathcal{T}_1^{\downarrow\downarrow}$ & $(10^{9}~\mathrm{Hz}/\mu_N)$
			& 4.15, 3.71		& [\onlinecite{SafSafRad13},p] \\
$\mathcal{T}_1^{\uparrow\uparrow}$ & $(10^{9}~\mathrm{Hz}/\mu_N)$
			& 2.42, 2.29		& [\onlinecite{SafSafRad13},p] \\
$\mathcal{T}_1^{\downarrow\uparrow}$ & $(10^{9}~\mathrm{Hz}/\mu_N)$
			& 1.13	& [p] \\
$\mathcal{T}_2^{\downarrow\downarrow}$ & $(10^{9}~\mathrm{Hz}/e\mathrm{b})$
			& 1.49, 1.47		& [\onlinecite{SafSafRad13},p] \\
$\mathcal{T}_2^{\uparrow\uparrow}$ & $(10^{9}~\mathrm{Hz}/e\mathrm{b})$
			& 1.67, 1.73		& [\onlinecite{SafSafRad13},p] \\
$\mathcal{T}_2^{\downarrow\uparrow}$ & $(10^{9}~\mathrm{Hz}/e\mathrm{b})$
			& 0.58		& [p] \\
$\mathcal{T}_3^{\downarrow\downarrow}$ & $(10^{3}~\mathrm{Hz}/\mu_N\mathrm{b})$
			& 15		& [p] \\
$\mathcal{T}_3^{\uparrow\uparrow}$ & $(10^{3}~\mathrm{Hz}/\mu_N\mathrm{b})$
			& $-12$	& [p] \\
$\mathcal{T}_4^{\downarrow\downarrow}$ & $(10^{3}~\mathrm{Hz}/e\mathrm{b}^2)$
			& 0.10	& [p] \\
$\mathcal{T}_4^{\uparrow\uparrow}$ & $(10^{3}~\mathrm{Hz}/e\mathrm{b}^2)$
			& 0.31	& [p] \\
$\Delta_\mathrm{fs}$ & $(10^{15}~\mathrm{Hz})$
			& 0.129682	& \cite{Cha58} \\
\end{tabular}
\footnotetext[1]{inferred from intrinsic quadrupole and hexadecapole moments}
\footnotetext[2]{from the theoretical value for $^{233}$U~\cite{Wil62}
}
\footnotetext[3]{see note \cite{grumpy}
}
\end{ruledtabular}
\end{table}

Estimating the terms $D_J$ and $D^\prime_J$ requires nuclear and electronic matrix elements, as well as the energy differences $\Delta_\mathrm{fs}$ and $\Delta_\mathrm{nuc}$. Table \ref{Tab:props} compiles the relevant properties, with values taken or inferred from the literature where available. Also included are present {\it ab initio} theoretical electronic matrix elements. The method starts by solving the self-consistent, fully-relativistic Dirac-Hartree-Fock (DHF) equations and includes important correlation corrections of the Brueckner orbital (BO) and random phase approximation (RPA) type in the calculation of matrix elements (see, e.g., Refs.~\cite{JohIdrSap87,Joh07}). For diagonal matrix elements, 
the present results may be compared with recent coupled-cluster results given by Safronova \etal{}~\cite{SafSafRad13}, which include a more extensive treatment of correlation effects. From Table \ref{Tab:props}, contributions to the $\widetilde{D}$ constants are estimated to be (in units of Hz) \cite{bashful}
\begin{equation}
\begin{gathered}
D_{5/2}\approx 0.6,
\qquad
D^\prime_{5/2}\approx 5+4\left(\frac{\eta_2}{0.15}\right)^2,
\\
D_{7/2}\approx 3,
\qquad
D^\prime_{7/2}\approx -36+8\left(\frac{\eta_2}{0.15}\right)^2,
\end{gathered}
\label{Eq:Destimates}
\end{equation}
where entries in Table \ref{Tab:props} yield values of $\eta_2$ spanning from $0.03$ to $0.15$. From the estimates given here, it is evident that higher order effects are non-negligible for the $\widetilde{D}$ constants. 
Residual terms $d^\prime_J$ 
are further estimated to be suppressed by more than an order-of-magnitude relative to the respective $D^\prime_J$.


In hypothetical absence of higher order effects, experimental $\widetilde{D}$ constants could be readily combined with theoretical matrix elements 
to extract the nuclear hexadecapole moment $\Pi\propto\mathcal{M}^{(4)}_{\downarrow\downarrow}$, similar to what has been done for nuclear dipole and quadrupole moments using $\widetilde{A}$ and $\widetilde{B}$ constants~\cite{CamRadKuz11,SafSafRad13}.
The availability of two constants, $\widetilde{D}_{5/2}$ and $\widetilde{D}_{7/2}$, would provide a degree of redundancy for this process. In the actual case---wherein second order effects are not absent or negligible---the ``extra'' constant provides an opportunity to suppress uncertainty in $\Pi$ resulting from these additional contributions. Hyperfine constants $B_{5/2}$ and $B_{7/2}$ appearing in Eq.~(\ref{Eq:Drecast}) may be determined to high precision with microwave spectroscopy (recall, $B=\widetilde{B}$ at the ppm level), while the off-diagonal constant $B_\mathrm{o.d.}$ can be expressed in terms of $B_{5/2}$ or $B_{7/2}$,
\begin{equation*}
B_\mathrm{o.d.}
=\frac{7}{12}
\left(\frac{\mathcal{T}^{(2)}_{\downarrow\uparrow}}{\mathcal{T}^{(2)}_{\downarrow\downarrow}}\right)
B_{5/2}
=\frac{5}{6\sqrt{2}}
\left(\frac{\mathcal{T}^{(2)}_{\downarrow\uparrow}}{\mathcal{T}^{(2)}_{\uparrow\uparrow}}\right)
B_{7/2},
\end{equation*}
such that evaluation of $B_\mathrm{o.d.}$ is limited by theoretical uncertainty in the ratio 
$\mathcal{T}^{(2)}_{\downarrow\uparrow}/
\mathcal{T}^{(2)}_{\downarrow\downarrow}$
or
$\mathcal{T}^{(2)}_{\downarrow\uparrow}/
\mathcal{T}^{(2)}_{\uparrow\uparrow}$.
Using coupled cluster techniques with empirical scaling, Safronova \etal~\cite{SafSafRad13} have demonstrated evaluation of the diagonal matrix elements $\mathcal{T}^{(2)}_{\downarrow\downarrow}$ and $\mathcal{T}^{(2)}_{\uparrow\uparrow}$ to $\sim\!1\%$, and a similar accuracy could be expected for the above ratios. Moreover, the two expressions for $B_\mathrm{o.d.}$ allow separate evaluations and further assessment of accuracy. With $B_{5/2}$, $B_{7/2}$, and $B_\mathrm{o.d.}$ known, an appropriate linear combination of $\widetilde{D}_{5/2}$ and $\widetilde{D}_{7/2}$ may be chosen to eliminate terms in Eq.~(\ref{Eq:Drecast}) proportional to the poorly-known factor $\eta_2^2$. Given theoretical values for $\mathcal{T}^{(4)}_{\downarrow\downarrow}$ and $\mathcal{T}^{(4)}_{\uparrow\uparrow}$, one may then solve the resulting expression for $\Pi\propto\mathcal{M}^{(4)}_{\downarrow\downarrow}$.

Perhaps a more intriguing prospect than obtaining $\Pi$ is the alternative: combining $\widetilde{D}_{5/2}$ and $\widetilde{D}_{7/2}$ to solve for $\eta_2$, as this parameter contains information about the nuclear isomer state, \thm{}. 
Taken in conjunction, Eqs.~(\ref{Eq:D3term}) and (\ref{Eq:Drecast}) yield an analytic solution for $\eta_2$ independent of the hexadecapole moment \cite{grumpy},
\begin{equation}
\begin{gathered}
\eta_2=
\sqrt{
2\frac{42875\Delta_\mathrm{fs}
X-108B_\mathrm{o.d.}^2(7\rho_D+1)}
{441B_{5/2}^2-700 \rho_D B_{7/2}^2+384B_\mathrm{o.d.}^2(7\rho_D-1)}
};
\\
X=\left(\widetilde{D}_{5/2}-d^\prime_{5/2}\right)
-\rho_D\left(\widetilde{D}_{7/2}-d^\prime_{7/2}\right),
\end{gathered}
\label{Eq:eta2}
\end{equation}
where $\rho_D\equiv D_{5/2}/D_{7/2}=\left(\sqrt{22}/7\right)
\left(\mathcal{T}^{(4)}_{\downarrow\downarrow}
/\mathcal{T}^{(4)}_{\uparrow\uparrow}\right)$.
Borrowing values from Table \ref{Tab:props}, we estimate that $\eta_2$ could potentially be determined to $\sim\!5\%$ using Eq.~(\ref{Eq:eta2}). To arrive at this conclusion, we ascribed plausible uncertainties to parameters on the right hand side of Eq.~(\ref{Eq:eta2}), assuming their precise evaluation with microwave spectroscopy and state-of-the-art theoretical techniques (e.g., Ref.~\cite{SafSafRad13}). Namely, we assumed uncertainties of $\sim\!1\%$ for $B_\mathrm{o.d.}$, $\sim\!20\%$ for $\rho_D$, and $\sim\!1\%$ for $(\widetilde{D}_J-d^\prime_J)$. Uncertainty propagation into $\eta_2$ was tracked by Monte Carlo evaluation of Eq.~(\ref{Eq:eta2}) with normally distributed parameters.
While $\sim\!5\%$ evaluation of $\eta_2$ is deemed a distinct possibility, we stress that accuracy at this level is not assured, even with the assumed uncertainties. For example, whereas a quasiparticle-plus-phonon model calculation predicts 2.4 $e$b for
$\mathcal{M}^{(2)}_{\downarrow\uparrow}$, a semi-empirical analysis predicts 0.80 $e$b (see Table~\ref{Tab:props}). Relative to the former, the latter value implies an order-of-magnitude reduction in the signal provided by $\eta_2^2$, with a corresponding reduction in the accuracy to which $\eta_2$ may be determined. Acknowledging the possibility of a larger $\mathcal{M}^{(2)}_{\downarrow\uparrow}$, on the other hand, suggests potentially better resolution of $\eta_2$. 

In analogy to the $\widetilde{D}$ constants, the $\widetilde{C}$ constants are likewise split into three terms,
\begin{equation}
\widetilde{C}_J=C_J+C^\prime_J+c^\prime_J,
\label{Eq:C3term}
\end{equation}
where $C_J$ is given by Eq.~(\ref{Eq:ABCD}), $C^\prime_J$ includes dominant second order contributions, and $c^\prime_J$ subsumes residual higher order contributions. For $\widetilde{C}$ constants, second order $M1$-$E2$ contributions emerge along with $E2$-$E2$ contributions, while $M1$-$M1$ contributions remain absent from angular considerations \cite{Sch55}.
Written analogously to Eq.~(\ref{Eq:Drecast}), the dominant second order contributions are
\begin{equation}
\begin{gathered}
\begin{aligned}
C_{5/2}^\prime=&
\frac{3}
{49000\,\Delta_\mathrm{fs}}
\bigl[
2100\sqrt{3}\,A_\mathrm{o.d.}B_\mathrm{o.d.}
-
792\,B_\mathrm{o.d.}^2
\\&
+
3675\sqrt{14}\,\eta_1\eta_2\,A_{5/2}B_{5/2}
+
147\,\eta_2^2\,B_{5/2}^2
\\&
-
700\sqrt{42}\,\eta_1\eta_2\,A_\mathrm{o.d.}B_\mathrm{o.d.}
-
352\,\eta_2^2\,B_\mathrm{o.d.}^2
\bigr]
,
\end{aligned}
\\
\begin{aligned}
C_{7/2}^\prime=&
\frac{1}
{7000\,\Delta_\mathrm{fs}}
\bigl[
-3150\sqrt{3}\,A_\mathrm{o.d.}B_\mathrm{o.d.}
-
324\,B_\mathrm{o.d.}^2
\\&
+
2625\sqrt{14}\,\eta_1\eta_2\,A_{7/2}B_{7/2}
+
50\,\eta_2^2\,B_{7/2}^2
\\&
-
1050\sqrt{42}\,\eta_1\eta_2\,A_\mathrm{o.d.}B_\mathrm{o.d.}
+
144\,\eta_2^2\,B_\mathrm{o.d.}^2
\bigr]
,
\end{aligned}
\end{gathered}
\label{Eq:Crecast}
\end{equation}
where we have introduced the off-diagonal constant 
$A_\mathrm{o.d.}\equiv (1/21)\sqrt{2/5}\,\mathcal{M}^{(1)}_{\downarrow\downarrow}
\mathcal{T}^{(1)}_{\downarrow\uparrow}$ and where $\eta_1$ is defined by Eq.~(\ref{Eq:etak}). From Table \ref{Tab:props}, contributions to the $\widetilde{C}$ constants are estimated to be (in units of Hz) \cite{bashful}
\begin{equation}
\begin{gathered}
C_{5/2}\approx 130,
\qquad
C_{7/2}\approx -120,
\\
C^\prime_{5/2}\approx 
-70
+60\left(\frac{\eta_1}{0.46}\right)\left(\frac{\eta_2}{0.15}\right)
+7\left(\frac{\eta_2}{0.15}\right)^2,
\\
C^\prime_{7/2}\approx 
-200
+20\left(\frac{\eta_1}{0.46}\right)\left(\frac{\eta_2}{0.15}\right)
+9\left(\frac{\eta_2}{0.15}\right)^2,
\end{gathered}
\label{Eq:Cestimates}
\end{equation}
where entries in Table \ref{Tab:props} yield values of $\left|\eta_1\right|$ spanning from $0.13$ to $0.46$. The hyperfine constants $A_{5/2}$ and $A_{7/2}$ in Eq.~(\ref{Eq:Crecast}) may be determined to high precision with microwave spectroscopy, while $A_\mathrm{o.d.}$ satisfies
\begin{equation*}
A_\mathrm{o.d.}
=\sqrt{\frac{5}{2}}
\left(\frac{\mathcal{T}^{(1)}_{\downarrow\uparrow}}{\mathcal{T}^{(1)}_{\downarrow\downarrow}}\right)
A_{5/2}
=\sqrt{6}
\left(\frac{\mathcal{T}^{(1)}_{\downarrow\uparrow}}{\mathcal{T}^{(1)}_{\uparrow\uparrow}}\right)
A_{7/2}.
\end{equation*}
Safronova \etal{}~\cite{SafSafRad13} have demonstrated $\sim\!1\%$ evaluation of $\mathcal{T}^{(1)}_{\downarrow\downarrow}$ and $\mathcal{T}^{(1)}_{\uparrow\uparrow}$, and a similar accuracy could be expected for the ratios appearing here. Equations (\ref{Eq:C3term}) and (\ref{Eq:Crecast}) may be combined to yield a solution for $\eta_1$ dependent upon $C_{5/2}$ and $C_{7/2}$ only through the ratio $\rho_C\equiv C_{5/2}/C_{7/2}=(1/7)\sqrt{110/3}\left(\mathcal{T}^{(3)}_{\downarrow\downarrow}/\mathcal{T}^{(3)}_{\uparrow\uparrow}\right)$ and, thus, independent of the nuclear octupole moment $\Omega\propto\mathcal{M}^{(3)}_{\downarrow\downarrow}$. The resulting analytic solution is lengthy and is not presented here. Borrowing values from Table \ref{Tab:props} and taking plausible uncertainties of $\sim\!1\%$ for $A_\mathrm{o.d.}$, $\sim\!20\%$ for $\rho_C$, and $\sim\!1\%$ for $(\widetilde{C}_J-c^\prime_J)$, assuming their evaluation with microwave spectroscopy and state-of-the-art theoretical techniques, we find that $\eta_1$ can potentially be determined to $\sim\!20\%$.

Zhao \etal{}~\cite{ZhaEscNat12} recently reported direct observation of the \thm{}\,$\rightarrow$\,\thg{} nuclear deexcitation in neutral thorium and concluded that the relaxation occurs via $M1$ decay with a lifetime $\tau=8.7\pm1.4$ hr, though we note that this result has been met with some skepticism (see Ref.~\cite{PeiZim13}).
The $M1$ decay rate depends on off-diagonal matrix element $\mathcal{M}^{(1)}_{\downarrow\uparrow}$, as well as $\Delta_\mathrm{nuc}$. Expressed in terms of $\eta_1$ in favor of $\mathcal{M}^{(1)}_{\downarrow\uparrow}$, this rate reads \cite{PorFla10,sleepy} $\tau^{-1}=(14/5)\hbar^{-4} c^{-3}\Delta_\mathrm{fs}^{-1}\Delta_\mathrm{nuc}^4\eta_1^2\mu^2$, where $\hbar$ is the reduced Planck constant and $c$ is the speed of light. 
Rearranging for $\Delta_\mathrm{nuc}$, this expression gives the proportionality relation
$
\Delta_\mathrm{nuc}\propto
\tau^{-1/4}\,\left|\eta_1\right|^{-1/2}\,\left|\mu\right|^{-1/2}.
$
With $\tau$ and $\mu$ given to $\sim\!15\%$ \cite{ZhaEscNat12} and $\sim\!1\%$~\cite{SafSafRad13}, respectively, and assuming $\eta_1$ to be evaluated to $\sim\!20\%$ following the above prescription, we conclude that $\Delta_\mathrm{nuc}$ can potentially be determined to $\sim\!10\%$. Thus, the present method has potential to obtain $\Delta_\mathrm{nuc}$ with accuracy comparable to the $\gamma$-ray differencing schemes of Refs.~\cite{HelRei94,BarArdBar03,GuiHel05,BecBecBei07,BecWuBei10}, which have reported inconsistent results, while using a completely independent approach to deducing this important property.

In summary, here we have identified a potential means to extract the \thg{}\,$\rightarrow$\,\thm{} nuclear excitation energy using a combination of precision microwave spectroscopy together with state-of-the-art theoretical methods of atomic structure. The triply-ionized species \thgion{} lends itself well to the present proposal. Firstly, the long-lived ground fine structure doublet $5F_{5/2,7/2}$ has a multitude of hyperfine intervals, with spectroscopic constants $\widetilde{C}_{5/2}$, $\widetilde{C}_{7/2}$, $\widetilde{D}_{5/2}$, and $\widetilde{D}_{7/2}$ having sizable fractional contributions attributed to hyperfine mixing with states of the nuclear isomer. Secondly, the single-valence character of \thgion{} greatly aids in its theoretical description (i.e., calculation of electronic properties), especially by comparison to the complex four-valence neutral system. Lastly, foundations of this proposal have already been demonstrated in recent works focused on this ion, including cooling, trapping, and spectroscopic interrogation within a Paul trap \cite{CamRadKuz11} and accurate calculation of electronic hyperfine matrix elements \cite{SafSafRad13}. The present work motivates continued effort in these directions as well as efforts towards improved characterization of the $M1$ decay rate of the isomer \cite{ZhaEscNat12,PeiZim13}. A new, independent determination of the nuclear excitation energy could prove to be an essential step in ultimately realizing direct laser-excitation of the thorium-229 nucleus.


\begin{thebibliography}{34}%
\makeatletter
\providecommand \@ifxundefined [1]{%
 \@ifx{#1\undefined}
}%
\providecommand \@ifnum [1]{%
 \ifnum #1\expandafter \@firstoftwo
 \else \expandafter \@secondoftwo
 \fi
}%
\providecommand \@ifx [1]{%
 \ifx #1\expandafter \@firstoftwo
 \else \expandafter \@secondoftwo
 \fi
}%
\providecommand \natexlab [1]{#1}%
\providecommand \enquote  [1]{``#1''}%
\providecommand \bibnamefont  [1]{#1}%
\providecommand \bibfnamefont [1]{#1}%
\providecommand \citenamefont [1]{#1}%
\providecommand \href@noop [0]{\@secondoftwo}%
\providecommand \href [0]{\begingroup \@sanitize@url \@href}%
\providecommand \@href[1]{\@@startlink{#1}\@@href}%
\providecommand \@@href[1]{\endgroup#1\@@endlink}%
\providecommand \@sanitize@url [0]{\catcode `\\12\catcode `\$12\catcode
  `\&12\catcode `\#12\catcode `\^12\catcode `\_12\catcode `\%12\relax}%
\providecommand \@@startlink[1]{}%
\providecommand \@@endlink[0]{}%
\providecommand \url  [0]{\begingroup\@sanitize@url \@url }%
\providecommand \@url [1]{\endgroup\@href {#1}{\urlprefix }}%
\providecommand \urlprefix  [0]{URL }%
\providecommand \Eprint [0]{\href }%
\providecommand \doibase [0]{http://dx.doi.org/}%
\providecommand \selectlanguage [0]{\@gobble}%
\providecommand \bibinfo  [0]{\@secondoftwo}%
\providecommand \bibfield  [0]{\@secondoftwo}%
\providecommand \translation [1]{[#1]}%
\providecommand \BibitemOpen [0]{}%
\providecommand \bibitemStop [0]{}%
\providecommand \bibitemNoStop [0]{.\EOS\space}%
\providecommand \EOS [0]{\spacefactor3000\relax}%
\providecommand \BibitemShut  [1]{\csname bibitem#1\endcsname}%
\let\auto@bib@innerbib\@empty
\bibitem [{\citenamefont {Peik}\ and\ \citenamefont {Tamm}(2003)}]{PeiTam03}%
  \BibitemOpen
  \bibfield  {author} {\bibinfo {author} {\bibfnamefont {E.}~\bibnamefont
  {Peik}}\ and\ \bibinfo {author} {\bibfnamefont {C.}~\bibnamefont {Tamm}},\
  }\href {http://stacks.iop.org/0295-5075/61/i=2/a=181} {\bibfield  {journal}
  {\bibinfo  {journal} {Europhys. Lett.}\ }\textbf {\bibinfo {volume} {61}},\
  \bibinfo {pages} {181} (\bibinfo {year} {2003})}\BibitemShut {NoStop}%
\bibitem [{\citenamefont {Campbell}\ \emph {et~al.}(2009)\citenamefont
  {Campbell}, \citenamefont {Steele}, \citenamefont {Churchill}, \citenamefont
  {DePalatis}, \citenamefont {Naylor}, \citenamefont {Matsukevich},
  \citenamefont {Kuzmich},\ and\ \citenamefont {Chapman}}]{CamSteChu09}%
  \BibitemOpen
  \bibfield  {author} {\bibinfo {author} {\bibfnamefont {C.~J.}\ \bibnamefont
  {Campbell}}, \bibinfo {author} {\bibfnamefont {A.~V.}\ \bibnamefont
  {Steele}}, \bibinfo {author} {\bibfnamefont {L.~R.}\ \bibnamefont
  {Churchill}}, \bibinfo {author} {\bibfnamefont {M.~V.}\ \bibnamefont
  {DePalatis}}, \bibinfo {author} {\bibfnamefont {D.~E.}\ \bibnamefont
  {Naylor}}, \bibinfo {author} {\bibfnamefont {D.~N.}\ \bibnamefont
  {Matsukevich}}, \bibinfo {author} {\bibfnamefont {A.}~\bibnamefont
  {Kuzmich}}, \ and\ \bibinfo {author} {\bibfnamefont {M.~S.}\ \bibnamefont
  {Chapman}},\ }\href {\doibase 10.1103/PhysRevLett.102.233004} {\bibfield
  {journal} {\bibinfo  {journal} {Phys. Rev. Lett.}\ }\textbf {\bibinfo
  {volume} {102}},\ \bibinfo {pages} {233004} (\bibinfo {year}
  {2009})}\BibitemShut {NoStop}%
\bibitem [{\citenamefont {Rellergert}\ \emph {et~al.}(2010)\citenamefont
  {Rellergert}, \citenamefont {DeMille}, \citenamefont {Greco}, \citenamefont
  {Hehlen}, \citenamefont {Torgerson},\ and\ \citenamefont
  {Hudson}}]{RelDeMGre10}%
  \BibitemOpen
  \bibfield  {author} {\bibinfo {author} {\bibfnamefont {W.~G.}\ \bibnamefont
  {Rellergert}}, \bibinfo {author} {\bibfnamefont {D.}~\bibnamefont {DeMille}},
  \bibinfo {author} {\bibfnamefont {R.~R.}\ \bibnamefont {Greco}}, \bibinfo
  {author} {\bibfnamefont {M.~P.}\ \bibnamefont {Hehlen}}, \bibinfo {author}
  {\bibfnamefont {J.~R.}\ \bibnamefont {Torgerson}}, \ and\ \bibinfo {author}
  {\bibfnamefont {E.~R.}\ \bibnamefont {Hudson}},\ }\href {\doibase
  10.1103/PhysRevLett.104.200802} {\bibfield  {journal} {\bibinfo  {journal}
  {Phys. Rev. Lett.}\ }\textbf {\bibinfo {volume} {104}},\ \bibinfo {pages}
  {200802} (\bibinfo {year} {2010})}\BibitemShut {NoStop}%
\bibitem [{\citenamefont {Campbell}\ \emph {et~al.}(2012)\citenamefont
  {Campbell}, \citenamefont {Radnaev}, \citenamefont {Kuzmich}, \citenamefont
  {Dzuba}, \citenamefont {Flambaum},\ and\ \citenamefont
  {Derevianko}}]{CamRadKuz12}%
  \BibitemOpen
  \bibfield  {author} {\bibinfo {author} {\bibfnamefont {C.~J.}\ \bibnamefont
  {Campbell}}, \bibinfo {author} {\bibfnamefont {A.~G.}\ \bibnamefont
  {Radnaev}}, \bibinfo {author} {\bibfnamefont {A.}~\bibnamefont {Kuzmich}},
  \bibinfo {author} {\bibfnamefont {V.~A.}\ \bibnamefont {Dzuba}}, \bibinfo
  {author} {\bibfnamefont {V.~V.}\ \bibnamefont {Flambaum}}, \ and\ \bibinfo
  {author} {\bibfnamefont {A.}~\bibnamefont {Derevianko}},\ }\href {\doibase
  10.1103/PhysRevLett.108.120802} {\bibfield  {journal} {\bibinfo  {journal}
  {Phys. Rev. Lett.}\ }\textbf {\bibinfo {volume} {108}},\ \bibinfo {pages}
  {120802} (\bibinfo {year} {2012})}\BibitemShut {NoStop}%
\bibitem [{\citenamefont {Kazakov}\ \emph {et~al.}(2012)\citenamefont
  {Kazakov}, \citenamefont {Litvinov}, \citenamefont {Romanenko}, \citenamefont
  {Yatsenko}, \citenamefont {Romanenko}, \citenamefont {Schreitl},
  \citenamefont {Winkler},\ and\ \citenamefont {Schumm}}]{KazLitRom12}%
  \BibitemOpen
  \bibfield  {author} {\bibinfo {author} {\bibfnamefont {G.~A.}\ \bibnamefont
  {Kazakov}}, \bibinfo {author} {\bibfnamefont {A.~N.}\ \bibnamefont
  {Litvinov}}, \bibinfo {author} {\bibfnamefont {V.~I.}\ \bibnamefont
  {Romanenko}}, \bibinfo {author} {\bibfnamefont {L.~P.}\ \bibnamefont
  {Yatsenko}}, \bibinfo {author} {\bibfnamefont {A.~V.}\ \bibnamefont
  {Romanenko}}, \bibinfo {author} {\bibfnamefont {M.}~\bibnamefont {Schreitl}},
  \bibinfo {author} {\bibfnamefont {G.}~\bibnamefont {Winkler}}, \ and\
  \bibinfo {author} {\bibfnamefont {T.}~\bibnamefont {Schumm}},\ }\href
  {http://stacks.iop.org/1367-2630/14/i=8/a=083019} {\bibfield  {journal}
  {\bibinfo  {journal} {New J. Phys.}\ }\textbf {\bibinfo {volume} {14}},\
  \bibinfo {pages} {083019} (\bibinfo {year} {2012})}\BibitemShut {NoStop}%
\bibitem [{\citenamefont {Flambaum}(2006)}]{Fla06}%
  \BibitemOpen
  \bibfield  {author} {\bibinfo {author} {\bibfnamefont {V.~V.}\ \bibnamefont
  {Flambaum}},\ }\href {\doibase 10.1103/PhysRevLett.97.092502} {\bibfield
  {journal} {\bibinfo  {journal} {Phys. Rev. Lett.}\ }\textbf {\bibinfo
  {volume} {97}},\ \bibinfo {pages} {092502} (\bibinfo {year}
  {2006})}\BibitemShut {NoStop}%
\bibitem [{\citenamefont {Tkalya}(2011)}]{Tka11}%
  \BibitemOpen
  \bibfield  {author} {\bibinfo {author} {\bibfnamefont {E.~V.}\ \bibnamefont
  {Tkalya}},\ }\href {\doibase 10.1103/PhysRevLett.106.162501} {\bibfield
  {journal} {\bibinfo  {journal} {Phys. Rev. Lett.}\ }\textbf {\bibinfo
  {volume} {106}},\ \bibinfo {pages} {162501} (\bibinfo {year}
  {2011})}\BibitemShut {NoStop}%
\bibitem [{\citenamefont {Reich}\ and\ \citenamefont
  {Helmer}(1990)}]{ReiHel90}%
  \BibitemOpen
  \bibfield  {author} {\bibinfo {author} {\bibfnamefont {C.~W.}\ \bibnamefont
  {Reich}}\ and\ \bibinfo {author} {\bibfnamefont {R.~G.}\ \bibnamefont
  {Helmer}},\ }\href {\doibase 10.1103/PhysRevLett.64.271} {\bibfield
  {journal} {\bibinfo  {journal} {Phys. Rev. Lett.}\ }\textbf {\bibinfo
  {volume} {64}},\ \bibinfo {pages} {271} (\bibinfo {year} {1990})}\BibitemShut
  {NoStop}%
\bibitem [{\citenamefont {Helmer}\ and\ \citenamefont
  {Reich}(1994)}]{HelRei94}%
  \BibitemOpen
  \bibfield  {author} {\bibinfo {author} {\bibfnamefont {R.~G.}\ \bibnamefont
  {Helmer}}\ and\ \bibinfo {author} {\bibfnamefont {C.~W.}\ \bibnamefont
  {Reich}},\ }\href {\doibase 10.1103/PhysRevC.49.1845} {\bibfield  {journal}
  {\bibinfo  {journal} {Phys. Rev. C}\ }\textbf {\bibinfo {volume} {49}},\
  \bibinfo {pages} {1845} (\bibinfo {year} {1994})}\BibitemShut {NoStop}%
\bibitem [{\citenamefont {Barci}\ \emph {et~al.}(2003)\citenamefont {Barci},
  \citenamefont {Ardisson}, \citenamefont {Barci-Funel}, \citenamefont {Weiss},
  \citenamefont {El~Samad},\ and\ \citenamefont {Sheline}}]{BarArdBar03}%
  \BibitemOpen
  \bibfield  {author} {\bibinfo {author} {\bibfnamefont {V.}~\bibnamefont
  {Barci}}, \bibinfo {author} {\bibfnamefont {G.}~\bibnamefont {Ardisson}},
  \bibinfo {author} {\bibfnamefont {G.}~\bibnamefont {Barci-Funel}}, \bibinfo
  {author} {\bibfnamefont {B.}~\bibnamefont {Weiss}}, \bibinfo {author}
  {\bibfnamefont {O.}~\bibnamefont {El~Samad}}, \ and\ \bibinfo {author}
  {\bibfnamefont {R.~K.}\ \bibnamefont {Sheline}},\ }\href {\doibase
  10.1103/PhysRevC.68.034329} {\bibfield  {journal} {\bibinfo  {journal} {Phys.
  Rev. C}\ }\textbf {\bibinfo {volume} {68}},\ \bibinfo {pages} {034329}
  (\bibinfo {year} {2003})}\BibitemShut {NoStop}%
\bibitem [{\citenamefont {{Guimar\~aes-Filho}}\ and\ \citenamefont
  {Helene}(2005)}]{GuiHel05}%
  \BibitemOpen
  \bibfield  {author} {\bibinfo {author} {\bibfnamefont {Z.~O.}\ \bibnamefont
  {{Guimar\~aes-Filho}}}\ and\ \bibinfo {author} {\bibfnamefont
  {O.}~\bibnamefont {Helene}},\ }\href {\doibase 10.1103/PhysRevC.71.044303}
  {\bibfield  {journal} {\bibinfo  {journal} {Phys. Rev. C}\ }\textbf {\bibinfo
  {volume} {71}},\ \bibinfo {pages} {044303} (\bibinfo {year}
  {2005})}\BibitemShut {NoStop}%
\bibitem [{\citenamefont {Beck}\ \emph {et~al.}(2007)\citenamefont {Beck},
  \citenamefont {Becker}, \citenamefont {Beiersdorfer}, \citenamefont {Brown},
  \citenamefont {Moody}, \citenamefont {Wilhelmy}, \citenamefont {Porter},
  \citenamefont {Kilbourne},\ and\ \citenamefont {Kelley}}]{BecBecBei07}%
  \BibitemOpen
  \bibfield  {author} {\bibinfo {author} {\bibfnamefont {B.~R.}\ \bibnamefont
  {Beck}}, \bibinfo {author} {\bibfnamefont {J.~A.}\ \bibnamefont {Becker}},
  \bibinfo {author} {\bibfnamefont {P.}~\bibnamefont {Beiersdorfer}}, \bibinfo
  {author} {\bibfnamefont {G.~V.}\ \bibnamefont {Brown}}, \bibinfo {author}
  {\bibfnamefont {K.~J.}\ \bibnamefont {Moody}}, \bibinfo {author}
  {\bibfnamefont {J.~B.}\ \bibnamefont {Wilhelmy}}, \bibinfo {author}
  {\bibfnamefont {F.~S.}\ \bibnamefont {Porter}}, \bibinfo {author}
  {\bibfnamefont {C.~A.}\ \bibnamefont {Kilbourne}}, \ and\ \bibinfo {author}
  {\bibfnamefont {R.~L.}\ \bibnamefont {Kelley}},\ }\href {\doibase
  10.1103/PhysRevLett.98.142501} {\bibfield  {journal} {\bibinfo  {journal}
  {Phys. Rev. Lett.}\ }\textbf {\bibinfo {volume} {98}},\ \bibinfo {pages}
  {142501} (\bibinfo {year} {2007})}\BibitemShut {NoStop}%
\bibitem [{\citenamefont {Beck}\ \emph {et~al.}()\citenamefont {Beck},
  \citenamefont {Wu}, \citenamefont {Beiersdorfer}, \citenamefont {Brown},
  \citenamefont {Becker}, \citenamefont {Moody}, \citenamefont {Wilhelmy},
  \citenamefont {Porter}, \citenamefont {Kilbourne},\ and\ \citenamefont
  {Kelley}}]{BecWuBei10}%
  \BibitemOpen
  \bibfield  {author} {\bibinfo {author} {\bibfnamefont {B.~R.}\ \bibnamefont
  {Beck}}, \bibinfo {author} {\bibfnamefont {C.}~\bibnamefont {Wu}}, \bibinfo
  {author} {\bibfnamefont {P.}~\bibnamefont {Beiersdorfer}}, \bibinfo {author}
  {\bibfnamefont {G.~V.}\ \bibnamefont {Brown}}, \bibinfo {author}
  {\bibfnamefont {J.~A.}\ \bibnamefont {Becker}}, \bibinfo {author}
  {\bibfnamefont {K.~J.}\ \bibnamefont {Moody}}, \bibinfo {author}
  {\bibfnamefont {J.~B.}\ \bibnamefont {Wilhelmy}}, \bibinfo {author}
  {\bibfnamefont {F.~S.}\ \bibnamefont {Porter}}, \bibinfo {author}
  {\bibfnamefont {C.~A.}\ \bibnamefont {Kilbourne}}, \ and\ \bibinfo {author}
  {\bibfnamefont {R.~L.}\ \bibnamefont {Kelley}},\ }\href@noop {} {}\bibinfo
  {note} {{in {\it Proceedings of the 12th International Conference on Nuclear
  Reaction Mechanisms}, edited by F. Cerutti and A. Ferrari (Varenna, Italy,
  2010), Vol. 1, pp. 255--258}}\BibitemShut {NoStop}%
\bibitem [{\citenamefont {Sakharov}(2010)}]{Sak10}%
  \BibitemOpen
  \bibfield  {author} {\bibinfo {author} {\bibfnamefont {S.}~\bibnamefont
  {Sakharov}},\ }\href {\doibase 10.1134/S1063778810010011} {\bibfield
  {journal} {\bibinfo  {journal} {Physics of Atomic Nuclei}\ }\textbf {\bibinfo
  {volume} {73}},\ \bibinfo {pages} {1} (\bibinfo {year} {2010})}\BibitemShut
  {NoStop}%
\bibitem [{\citenamefont {Campbell}\ \emph {et~al.}(2011)\citenamefont
  {Campbell}, \citenamefont {Radnaev},\ and\ \citenamefont
  {Kuzmich}}]{CamRadKuz11}%
  \BibitemOpen
  \bibfield  {author} {\bibinfo {author} {\bibfnamefont {C.~J.}\ \bibnamefont
  {Campbell}}, \bibinfo {author} {\bibfnamefont {A.~G.}\ \bibnamefont
  {Radnaev}}, \ and\ \bibinfo {author} {\bibfnamefont {A.}~\bibnamefont
  {Kuzmich}},\ }\href {\doibase 10.1103/PhysRevLett.106.223001} {\bibfield
  {journal} {\bibinfo  {journal} {Phys. Rev. Lett.}\ }\textbf {\bibinfo
  {volume} {106}},\ \bibinfo {pages} {223001} (\bibinfo {year}
  {2011})}\BibitemShut {NoStop}%
\bibitem [{\citenamefont {Hehlen}\ \emph {et~al.}(2013)\citenamefont {Hehlen},
  \citenamefont {Greco}, \citenamefont {Rellergert}, \citenamefont {Sullivan},
  \citenamefont {DeMille}, \citenamefont {Jackson}, \citenamefont {Hudson},\
  and\ \citenamefont {Torgerson}}]{HehGreRel13}%
  \BibitemOpen
  \bibfield  {author} {\bibinfo {author} {\bibfnamefont {M.~P.}\ \bibnamefont
  {Hehlen}}, \bibinfo {author} {\bibfnamefont {R.~R.}\ \bibnamefont {Greco}},
  \bibinfo {author} {\bibfnamefont {W.~G.}\ \bibnamefont {Rellergert}},
  \bibinfo {author} {\bibfnamefont {S.~T.}\ \bibnamefont {Sullivan}}, \bibinfo
  {author} {\bibfnamefont {D.}~\bibnamefont {DeMille}}, \bibinfo {author}
  {\bibfnamefont {R.~A.}\ \bibnamefont {Jackson}}, \bibinfo {author}
  {\bibfnamefont {E.~R.}\ \bibnamefont {Hudson}}, \ and\ \bibinfo {author}
  {\bibfnamefont {J.~R.}\ \bibnamefont {Torgerson}},\ }\href {\doibase
  http://dx.doi.org/10.1016/j.jlumin.2011.09.037} {\bibfield  {journal}
  {\bibinfo  {journal} {Journal of Luminescence}\ }\textbf {\bibinfo {volume}
  {133}},\ \bibinfo {pages} {91 } (\bibinfo {year} {2013})}\BibitemShut
  {NoStop}%
\bibitem [{\citenamefont {Safronova}\ \emph {et~al.}()\citenamefont
  {Safronova}, \citenamefont {Safronova}, \citenamefont {Radnaev},
  \citenamefont {Campbell},\ and\ \citenamefont {Kuzmich}}]{SafSafRad13}%
  \BibitemOpen
  \bibfield  {author} {\bibinfo {author} {\bibfnamefont {M.~S.}\ \bibnamefont
  {Safronova}}, \bibinfo {author} {\bibfnamefont {U.~I.}\ \bibnamefont
  {Safronova}}, \bibinfo {author} {\bibfnamefont {A.~G.}\ \bibnamefont
  {Radnaev}}, \bibinfo {author} {\bibfnamefont {C.~J.}\ \bibnamefont
  {Campbell}}, \ and\ \bibinfo {author} {\bibfnamefont {A.}~\bibnamefont
  {Kuzmich}},\ }\href@noop {} {}\bibinfo {note} {{arXiv:1305.0667v1
  (2013)}}\BibitemShut {NoStop}%
\bibitem [{\citenamefont {Zhao}\ \emph {et~al.}(2012)\citenamefont {Zhao},
  \citenamefont {Martinez~de Escobar}, \citenamefont {Rundberg}, \citenamefont
  {Bond}, \citenamefont {Moody},\ and\ \citenamefont {Vieira}}]{ZhaEscNat12}%
  \BibitemOpen
  \bibfield  {author} {\bibinfo {author} {\bibfnamefont {X.}~\bibnamefont
  {Zhao}}, \bibinfo {author} {\bibfnamefont {Y.~N.}\ \bibnamefont {Martinez~de
  Escobar}}, \bibinfo {author} {\bibfnamefont {R.}~\bibnamefont {Rundberg}},
  \bibinfo {author} {\bibfnamefont {E.~M.}\ \bibnamefont {Bond}}, \bibinfo
  {author} {\bibfnamefont {A.}~\bibnamefont {Moody}}, \ and\ \bibinfo {author}
  {\bibfnamefont {D.~J.}\ \bibnamefont {Vieira}},\ }\href {\doibase
  10.1103/PhysRevLett.109.160801} {\bibfield  {journal} {\bibinfo  {journal}
  {Phys. Rev. Lett.}\ }\textbf {\bibinfo {volume} {109}},\ \bibinfo {pages}
  {160801} (\bibinfo {year} {2012})}\BibitemShut {NoStop}%
\bibitem [{\citenamefont {Schwartz}(1955)}]{Sch55}%
  \BibitemOpen
  \bibfield  {author} {\bibinfo {author} {\bibfnamefont {C.}~\bibnamefont
  {Schwartz}},\ }\href {\doibase 10.1103/PhysRev.97.380} {\bibfield  {journal}
  {\bibinfo  {journal} {Phys. Rev.}\ }\textbf {\bibinfo {volume} {97}},\
  \bibinfo {pages} {380} (\bibinfo {year} {1955})}\BibitemShut {NoStop}%
\bibitem [{\citenamefont {Johnson}(2007)}]{Joh07}%
  \BibitemOpen
  \bibfield  {author} {\bibinfo {author} {\bibfnamefont {W.~R.}\ \bibnamefont
  {Johnson}},\ }\href@noop {} {\emph {\bibinfo {title} {Atomic Structure
  Theory: Lectures on Atomic Physics}}}\ (\bibinfo  {publisher} {Springer},\
  \bibinfo {address} {New York, NY},\ \bibinfo {year} {2007})\BibitemShut
  {NoStop}%
\bibitem [{\citenamefont {Beloy}\ \emph {et~al.}(2008)\citenamefont {Beloy},
  \citenamefont {Derevianko},\ and\ \citenamefont {Johnson}}]{BelDerJoh08}%
  \BibitemOpen
  \bibfield  {author} {\bibinfo {author} {\bibfnamefont {K.}~\bibnamefont
  {Beloy}}, \bibinfo {author} {\bibfnamefont {A.}~\bibnamefont {Derevianko}}, \
  and\ \bibinfo {author} {\bibfnamefont {W.~R.}\ \bibnamefont {Johnson}},\
  }\href {\doibase 10.1103/PhysRevA.77.012512} {\bibfield  {journal} {\bibinfo
  {journal} {Phys. Rev. A}\ }\textbf {\bibinfo {volume} {77}},\ \bibinfo
  {pages} {012512} (\bibinfo {year} {2008})}\BibitemShut {NoStop}%
\bibitem [{\citenamefont {Porsev}\ and\ \citenamefont
  {Flambaum}(2010)}]{PorFla10}%
  \BibitemOpen
  \bibfield  {author} {\bibinfo {author} {\bibfnamefont {S.~G.}\ \bibnamefont
  {Porsev}}\ and\ \bibinfo {author} {\bibfnamefont {V.~V.}\ \bibnamefont
  {Flambaum}},\ }\href {\doibase 10.1103/PhysRevA.81.032504} {\bibfield
  {journal} {\bibinfo  {journal} {Phys. Rev. A}\ }\textbf {\bibinfo {volume}
  {81}},\ \bibinfo {pages} {032504} (\bibinfo {year} {2010})}\BibitemShut
  {NoStop}%
\bibitem [{\citenamefont {Gerstenkorn}\ \emph {et~al.}(1974)\citenamefont
  {Gerstenkorn}, \citenamefont {Luc}, \citenamefont {Verges}, \citenamefont
  {Englekemeir}, \citenamefont {Gindler},\ and\ \citenamefont
  {Tomkins}}]{GerLucVer74}%
  \BibitemOpen
  \bibfield  {author} {\bibinfo {author} {\bibfnamefont {S.}~\bibnamefont
  {Gerstenkorn}}, \bibinfo {author} {\bibfnamefont {P.}~\bibnamefont {Luc}},
  \bibinfo {author} {\bibfnamefont {J.}~\bibnamefont {Verges}}, \bibinfo
  {author} {\bibfnamefont {D.}~\bibnamefont {Englekemeir}}, \bibinfo {author}
  {\bibfnamefont {J.}~\bibnamefont {Gindler}}, \ and\ \bibinfo {author}
  {\bibfnamefont {F.}~\bibnamefont {Tomkins}},\ }\href
  {http://www.osti.gov/scitech/servlets/purl/4237974} {\bibfield  {journal}
  {\bibinfo  {journal} {J. Phys. (Paris)}\ }\textbf {\bibinfo {volume} {35}},\
  \bibinfo {pages} {483} (\bibinfo {year} {1974})}\BibitemShut {NoStop}%
\bibitem [{\citenamefont {Bemis}\ \emph {et~al.}(1988)\citenamefont {Bemis},
  \citenamefont {McGowan}, \citenamefont {Jr}, \citenamefont {Milner},
  \citenamefont {Robinson}, \citenamefont {Stelson}, \citenamefont {Leander},\
  and\ \citenamefont {Reich}}]{BemMcGFor88}%
  \BibitemOpen
  \bibfield  {author} {\bibinfo {author} {\bibfnamefont {C.~E.}\ \bibnamefont
  {Bemis}}, \bibinfo {author} {\bibfnamefont {F.~K.}\ \bibnamefont {McGowan}},
  \bibinfo {author} {\bibfnamefont {J.~L. C.~F.}\ \bibnamefont {Jr}}, \bibinfo
  {author} {\bibfnamefont {W.~T.}\ \bibnamefont {Milner}}, \bibinfo {author}
  {\bibfnamefont {R.~L.}\ \bibnamefont {Robinson}}, \bibinfo {author}
  {\bibfnamefont {P.~H.}\ \bibnamefont {Stelson}}, \bibinfo {author}
  {\bibfnamefont {G.~A.}\ \bibnamefont {Leander}}, \ and\ \bibinfo {author}
  {\bibfnamefont {C.~W.}\ \bibnamefont {Reich}},\ }\href
  {http://stacks.iop.org/1402-4896/38/i=5/a=004} {\bibfield  {journal}
  {\bibinfo  {journal} {Phys. Sc.}\ }\textbf {\bibinfo {volume} {38}},\
  \bibinfo {pages} {657} (\bibinfo {year} {1988})}\BibitemShut {NoStop}%
\bibitem [{\citenamefont {Dykhne}\ and\ \citenamefont
  {Tkalya}(1998)}]{DykTka98}%
  \BibitemOpen
  \bibfield  {author} {\bibinfo {author} {\bibfnamefont {A.}~\bibnamefont
  {Dykhne}}\ and\ \bibinfo {author} {\bibfnamefont {E.}~\bibnamefont
  {Tkalya}},\ }\href {\doibase 10.1134/1.567659} {\bibfield  {journal}
  {\bibinfo  {journal} {JETP}\ }\textbf {\bibinfo {volume} {67}},\ \bibinfo
  {pages} {251} (\bibinfo {year} {1998})}\BibitemShut {NoStop}%
\bibitem [{\citenamefont {Ruchowska}\ \emph {et~al.}(2006)\citenamefont
  {Ruchowska}, \citenamefont {P\l{}\'ociennik}, \citenamefont
  {\ifmmode~\dot{Z}\else \.{Z}\fi{}ylicz}, \citenamefont {Mach}, \citenamefont
  {Kvasil}, \citenamefont {Algora}, \citenamefont {Amzal}, \citenamefont
  {B\"ack}, \citenamefont {Borge}, \citenamefont {Boutami}, \citenamefont
  {Butler}, \citenamefont {Cederk\"all}, \citenamefont {Cederwall},
  \citenamefont {Fogelberg}, \citenamefont {Fraile}, \citenamefont {Fynbo},
  \citenamefont {Hageb\o{}}, \citenamefont {Hoff}, \citenamefont {Gausemel},
  \citenamefont {Jungclaus}, \citenamefont {Kaczarowski}, \citenamefont
  {Kerek}, \citenamefont {Kurcewicz}, \citenamefont {Lagergren}, \citenamefont
  {Nacher}, \citenamefont {Rubio}, \citenamefont {Syntfeld}, \citenamefont
  {Tengblad}, \citenamefont {Wasilewski},\ and\ \citenamefont
  {Weissman}}]{RucPloZyl06}%
  \BibitemOpen
  \bibfield  {author} {\bibinfo {author} {\bibfnamefont {E.}~\bibnamefont
  {Ruchowska}}, \bibinfo {author} {\bibfnamefont {W.~A.}\ \bibnamefont
  {P\l{}\'ociennik}}, \bibinfo {author} {\bibfnamefont {J.}~\bibnamefont
  {\ifmmode~\dot{Z}\else \.{Z}\fi{}ylicz}}, \bibinfo {author} {\bibfnamefont
  {H.}~\bibnamefont {Mach}}, \bibinfo {author} {\bibfnamefont {J.}~\bibnamefont
  {Kvasil}}, \bibinfo {author} {\bibfnamefont {A.}~\bibnamefont {Algora}},
  \bibinfo {author} {\bibfnamefont {N.}~\bibnamefont {Amzal}}, \bibinfo
  {author} {\bibfnamefont {T.}~\bibnamefont {B\"ack}}, \bibinfo {author}
  {\bibfnamefont {M.~G.}\ \bibnamefont {Borge}}, \bibinfo {author}
  {\bibfnamefont {R.}~\bibnamefont {Boutami}}, \bibinfo {author} {\bibfnamefont
  {P.~A.}\ \bibnamefont {Butler}}, \bibinfo {author} {\bibfnamefont
  {J.}~\bibnamefont {Cederk\"all}}, \bibinfo {author} {\bibfnamefont
  {B.}~\bibnamefont {Cederwall}}, \bibinfo {author} {\bibfnamefont
  {B.}~\bibnamefont {Fogelberg}}, \bibinfo {author} {\bibfnamefont {L.~M.}\
  \bibnamefont {Fraile}}, \bibinfo {author} {\bibfnamefont {H.~O.~U.}\
  \bibnamefont {Fynbo}}, \bibinfo {author} {\bibfnamefont {E.}~\bibnamefont
  {Hageb\o{}}}, \bibinfo {author} {\bibfnamefont {P.}~\bibnamefont {Hoff}},
  \bibinfo {author} {\bibfnamefont {H.}~\bibnamefont {Gausemel}}, \bibinfo
  {author} {\bibfnamefont {A.}~\bibnamefont {Jungclaus}}, \bibinfo {author}
  {\bibfnamefont {R.}~\bibnamefont {Kaczarowski}}, \bibinfo {author}
  {\bibfnamefont {A.}~\bibnamefont {Kerek}}, \bibinfo {author} {\bibfnamefont
  {W.}~\bibnamefont {Kurcewicz}}, \bibinfo {author} {\bibfnamefont
  {K.}~\bibnamefont {Lagergren}}, \bibinfo {author} {\bibfnamefont
  {E.}~\bibnamefont {Nacher}}, \bibinfo {author} {\bibfnamefont
  {B.}~\bibnamefont {Rubio}}, \bibinfo {author} {\bibfnamefont
  {A.}~\bibnamefont {Syntfeld}}, \bibinfo {author} {\bibfnamefont
  {O.}~\bibnamefont {Tengblad}}, \bibinfo {author} {\bibfnamefont {A.~A.}\
  \bibnamefont {Wasilewski}}, \ and\ \bibinfo {author} {\bibfnamefont
  {L.}~\bibnamefont {Weissman}},\ }\href {\doibase 10.1103/PhysRevC.73.044326}
  {\bibfield  {journal} {\bibinfo  {journal} {Phys. Rev. C}\ }\textbf {\bibinfo
  {volume} {73}},\ \bibinfo {pages} {044326} (\bibinfo {year}
  {2006})}\BibitemShut {NoStop}%
\bibitem [{\citenamefont {Charles}()}]{Cha58}%
  \BibitemOpen
  \bibfield  {author} {\bibinfo {author} {\bibfnamefont {G.~W.}\ \bibnamefont
  {Charles}},\ }\href@noop {} {}\bibinfo {note} {{Oak Ridge National Laboratory
  Technical Report No. ORNL-2319, 1958 (unpublished)}}\BibitemShut {NoStop}%
\bibitem [{\citenamefont {Williams}(1962)}]{Wil62}%
  \BibitemOpen
  \bibfield  {author} {\bibinfo {author} {\bibfnamefont {S.~A.}\ \bibnamefont
  {Williams}},\ }\href {\doibase 10.1103/PhysRev.125.340} {\bibfield  {journal}
  {\bibinfo  {journal} {Phys. Rev.}\ }\textbf {\bibinfo {volume} {125}},\
  \bibinfo {pages} {340} (\bibinfo {year} {1962})}\BibitemShut {NoStop}%
\bibitem [{gru()}]{grumpy}%
  \BibitemOpen
  \href@noop {} {}\bibinfo {note} {{Throughout, off-diagonal matrix elements
  $\mathcal{T}^{(k)}_{\downarrow\uparrow}$ and
  $\mathcal{M}^{(k)}_{\downarrow\uparrow}$ are taken real for all $k$
  \cite{Llo51}. We arbitrarily fix $\mathcal{M}^{(2)}_{\downarrow\uparrow}$ to
  be positive, with one consequence being that $\eta_2$ is
  positive.}}\BibitemShut {Stop}%
\bibitem [{\citenamefont {Johnson}\ \emph {et~al.}(1987)\citenamefont
  {Johnson}, \citenamefont {Idrees},\ and\ \citenamefont
  {Sapirstein}}]{JohIdrSap87}%
  \BibitemOpen
  \bibfield  {author} {\bibinfo {author} {\bibfnamefont {W.~R.}\ \bibnamefont
  {Johnson}}, \bibinfo {author} {\bibfnamefont {M.}~\bibnamefont {Idrees}}, \
  and\ \bibinfo {author} {\bibfnamefont {J.}~\bibnamefont {Sapirstein}},\
  }\href {\doibase 10.1103/PhysRevA.35.3218} {\bibfield  {journal} {\bibinfo
  {journal} {Phys. Rev. A}\ }\textbf {\bibinfo {volume} {35}},\ \bibinfo
  {pages} {3218} (\bibinfo {year} {1987})}\BibitemShut {NoStop}%
\bibitem [{bas()}]{bashful}%
  \BibitemOpen
  \href@noop {} {}\bibinfo {note} {{The $5F_{5/2,7/2}$ hyperfine sub-levels are
  estimated to have widths $\Gamma(5F_{5/2})\ll\Gamma(5F_{7/2})\sim
  0.1~\mathrm{Hz}$ \cite{Joh07}, with each containing a magnetic
  field-insensitive ($M=0$) state. Supposing measurement of the hyperfine
  intervals to ${\sim\!0.1}$ Hz implies specification of the $\widetilde{C}$
  constants to ${\sim\!10^{-2}}$ Hz and the $\widetilde{D}$ constants to
  ${\sim\!10^{-3}}$ Hz.}}\BibitemShut {Stop}%
\bibitem [{\citenamefont {Peik}\ and\ \citenamefont
  {Zimmermann}(2013)}]{PeiZim13}%
  \BibitemOpen
  \bibfield  {author} {\bibinfo {author} {\bibfnamefont {E.}~\bibnamefont
  {Peik}}\ and\ \bibinfo {author} {\bibfnamefont {K.}~\bibnamefont
  {Zimmermann}},\ }\href {\doibase 10.1103/PhysRevLett.111.018901} {\bibfield
  {journal} {\bibinfo  {journal} {Phys. Rev. Lett.}\ }\textbf {\bibinfo
  {volume} {111}},\ \bibinfo {pages} {018901} (\bibinfo {year}
  {2013})}\BibitemShut {NoStop}%
\bibitem [{sle()}]{sleepy}%
  \BibitemOpen
  \href@noop {} {}\bibinfo {note} {{We assume Gaussian electromagnetic
  expressions such that $\mu_N^2=(e\hbar/2m_pc)^2=(\alpha/4)(\hbar c)^3(m_p
  c^2)^{-2}$, with $\alpha$ the fine structure constant and $m_p$ the proton
  mass.}}\BibitemShut {Stop}%
\bibitem [{\citenamefont {Lloyd}(1951)}]{Llo51}%
  \BibitemOpen
  \bibfield  {author} {\bibinfo {author} {\bibfnamefont {S.~P.}\ \bibnamefont
  {Lloyd}},\ }\href {\doibase 10.1103/PhysRev.81.161.2} {\bibfield  {journal}
  {\bibinfo  {journal} {Phys. Rev.}\ }\textbf {\bibinfo {volume} {81}},\
  \bibinfo {pages} {161} (\bibinfo {year} {1951})}\BibitemShut {NoStop}%
\end{thebibliography}

%

\end{document}